\documentclass[twocolumn,superscriptaddress,showpacs,preprintnumbers,amsmath,amssymb,prc]{revtex4}

\usepackage[dvipdfm]{graphicx}
\usepackage{dcolumn}
\usepackage{bm}
\usepackage{longtable}

\usepackage[dvips]{color} 

{%
   \begin{list}{\parbox{1zw}{$\bullet$}}
   {%
      \setlength{\itemsep}{0em}
      \setlength{\parsep}{0em}
   }
}{%
   \end{list}%
}

\begin{document}


\title{
Shallow and diffuse
spin-orbit potential for proton elastic scattering \\
from neutron-rich helium
isotopes at 71~MeV/nucleon}

\author{S.~Sakaguchi}
 \altaffiliation{Present address: Department of Physics, Kyushu
 University, Fukuoka 812-8581, Japan;\\ Electronic address:
 {\tt sakaguchi@phys.kyushu-u.ac.jp}}
\affiliation{Center for Nuclear Study, University of Tokyo,
 Tokyo 113-0001, Japan}
\author{T.~Uesaka}
\affiliation{RIKEN Nishina Center, Saitama 351-0198, Japan}
\author{N.~Aoi}
\affiliation{RIKEN Nishina Center, Saitama 351-0198, Japan}
\author{Y.~Ichikawa}
\affiliation{Department of Physics, University of Tokyo, Tokyo 113-0033, Japan}
\author{K.~Itoh}
\affiliation{Department of Physics, Saitama University, Saitama 338-8570, Japan}
\author{M.~Itoh}
\affiliation{Cyclotron \& Radioisotope Center, Tohoku University, Miyagi 980-8578, Japan}
\author{T.~Kawabata}
\affiliation{Center for Nuclear Study, University of Tokyo,
 Tokyo 113-0001, Japan}
\author{T.~Kawahara}
\affiliation{Department of Physics, Toho University, Chiba, Japan}
\author{Y.~Kondo}
\affiliation{Department of Physics, Tokyo Institute of Technology, Tokyo 152-8551, Japan}
\author{H.~Kuboki}
\affiliation{Department of Physics, University of Tokyo, Tokyo 113-0033, Japan}
\author{T.~Nakamura}
\affiliation{Department of Physics, Tokyo Institute of Technology, Tokyo 152-8551, Japan}
\author{T.~Nakao}
\affiliation{Department of Physics, University of Tokyo, Tokyo 113-0033, Japan}
\author{Y.~Nakayama}
\affiliation{Department of Physics, Tokyo Institute of Technology, Tokyo 152-8551, Japan}
\author{H.~Sakai}
\affiliation{Department of Physics, University of Tokyo, Tokyo 113-0033, Japan}
\author{Y.~Sasamoto}
\affiliation{Center for Nuclear Study, University of Tokyo,
 Tokyo 113-0001, Japan}
\author{K.~Sekiguchi}
\affiliation{RIKEN Nishina Center, Saitama 351-0198, Japan}
\author{T.~Shimamura}
\affiliation{Department of Physics, Tokyo Institute of Technology, Tokyo 152-8551, Japan}
\author{Y.~Shimizu}
\affiliation{Center for Nuclear Study, University of Tokyo,
 Tokyo 113-0001, Japan}
\author{T.~Wakui}
\affiliation{Cyclotron \& Radioisotope Center, Tohoku University, Miyagi 980-8578, Japan}
\date{\today}

\begin{abstract}
Vector analyzing powers for proton elastic scattering from $^8$He at
71~MeV/nucleon have been measured using a solid polarized proton
target operated in a low magnetic field of 0.1~T. The spin-orbit
potential obtained from a phenomenological optical model analysis is
found to be significantly shallower and more diffuse than the global
systematics of stable nuclei, which is an indication that the
spin-orbit potential is modified for scattering involving
neutron-rich nuclei. A close similarity between the matter radius
and the root-mean-square radius of the spin-orbit potential is also
identified.
\end{abstract}


\maketitle

The strong spin-orbit coupling in atomic nuclei plays an important
role in nuclear structure and reactions. One good example is the
spin-orbit splitting of single-particle levels, which is a key
ingredient for the success of the nuclear shell
model~\cite{Mayer49,Jensen49}. Spin-orbit coupling is also
responsible for many other phenomena such as the dominance of the
prolate shape and the emergence of the isomeric intruder state.
Moreover, in terms of nuclear reactions, spin-orbit coupling is
responsible for the polarization effects in elastic scattering.
There has recently been renewed interest in spin-orbit coupling
since it is predicted to be modified in neutron-rich nuclei. A
number of experimental results suggest a change in the shell
structure of neutron-rich nuclei that could be explained by a
reduction in the spin-orbit
splitting~\cite{Dobaczewski94,Lalazissis98,Otsuka06}. However, there
has been no experimental study examining how the spin-orbit
coupling is modified in nuclear reactions of unstable nuclei.

Spin asymmetry in proton--nucleus ($p$--$A$) elastic scattering is a
prominent manifestation of the spin-orbit coupling in nuclear
reactions. The coupling is generally represented by a spin-orbit
term in the optical model potential, i.e., the spin-orbit
potential. Current understanding of this potential has been based on
extensive measurements and analysis of the vector analyzing powers
for elastic scattering of polarized protons from various stable
nuclei over a wide energy range
~\cite{Craig64,Blumberg66,Comparat74,Sakaguchi82,Varner91,Koning03}.
It is now well established that the shape and magnitude of the
spin-orbit potential does not depend strongly on the target
nucleus. The shape is reasonably expressed by a derivative of the
density distribution~\cite{Dover72,Scheerbaum76,Bauge98}, while the
magnitude is almost independent of the mass
number~\cite{Varner91,Koning03}. However, whether these systematics
hold even in regions far from the stability line is still an open
question. The structure of neutron-rich nuclei often shows
distinctive features such as a very diffuse nuclear surface, a
neutron-skin and halo, and a difference between the radial
dependence of the proton and neutron distributions. From the surface
nature of the spin-orbit coupling, we can expect that the
spin-orbit potential is modified in the neutron-rich region. In
this letter we determine the spin-orbit potential between a proton
and a typical neutron-rich nucleus $^8$He and investigate the effect
of the exotic structure of the neutron-rich nucleus on the
spin-orbit coupling in $p$--$A$ scattering.

Determination of the spin-orbit potential requires vector analyzing
power data, and until several years ago, 
such data could not be obtained 
in the experiment with a radioactive-ion beam.
This was due to the lack of polarized targets that can be
operated at a low magnetic field of $\ll$1~T. However, we were
able to construct a solid polarized proton target at 0.1~T based on a new
polarizing method~\cite{Uesaka04,Wakui05,Hatano05,WakuiPST} and have
applied it to scattering experiments of $^6$He at
71~MeV/nucleon~\cite{Hatano05,Uesaka10,Sakaguchi11}.

Recently, we have measured the vector analyzing powers for proton
elastic scattering from $^8$He at 71~MeV/nucleon. These neutron-rich
helium isotopes are suitable for exploring the modification of spin-orbit
potential, since they have large neutron-excess ratios $(N-Z)/A$ and
significantly diffuse density distributions. The data are analyzed
with a phenomenological optical model to discuss the overall
characteristics of the spin-orbit interaction with a least-biased
approach. Details of both the measurements and analysis are reported
in this paper.

The analyzing power measurement of $p$--$^8$He elastic scattering
was carried out at RI Beam Factory operated by RIKEN
Nishina Center and Center for Nuclear Study, University of Tokyo.
The $^8$He beam was produced by a projectile fragmentation reaction
of an $^{18}$O beam with an energy of 100~MeV/nucleon bombarding a
13-mm thick Be target. The $^8$He particles were then separated by
the RIKEN Projectile-fragment Separator (RIPS)~\cite{Kubo92}. The
energy of the $^8$He beam was 71.0 $\pm$ 1.4~MeV/nucleon at the
center of the secondary target. The typical intensity and purity of
the beam were 1.5 $\times$ 10$^5$~pps and 77 \%, respectively. As a
secondary target, the solid polarized proton
target~\cite{Uesaka04,Wakui05,Hatano05,WakuiPST} was placed at the
final focal plane of the RIPS. The target was operated under a low
magnetic field of 91~mT, which allowed us to detect low-energy
($\sim$10 MeV) recoil protons under inverse kinematics conditions.
The average target polarization was 11.3 $\pm$ 2.6 \%.

The detector system is same as that used in the $p$--$^6$He
measurement described in Ref.~\cite{Sakaguchi11} except for the
recoil proton detectors. To achieve higher angular and energy
resolutions for the recoil protons, we used multi-wire drift
chambers (MWDCs) and CsI(Tl) scintillators with a Si PIN photodiode
readout. The position resolution of the MWDCs was 200~$\mu$m (full
width half maximum). This corresponds to an angular resolution of
$0.05^{\circ}$ in sigma in the center-of-mass system, which is one
order of magnitude better than that in the $p$--$^6$He measurement.
The effects of the magnetic field on the proton scattering angle,
which was comparable to or smaller than the detector resolution,
were properly corrected in the data analysis. Using the correlation
between the recoil and scattered particle scattering angles, a clear
peak corresponding to the $p$--$^8$He elastic scattering was
identified. Spurious asymmetries such as imbalances in the detector
efficiency and solid angle were canceled out by reversing the
direction of target polarization. It should be emphasized again that
the operation of the polarized target in a low magnetic field
allowed us to detect recoil protons with an angular resolution
sufficient to identify 
the elastic scattering 
events.

\begin{figure}[ht]
\centering
\resizebox{0.8\linewidth}{!}{\includegraphics{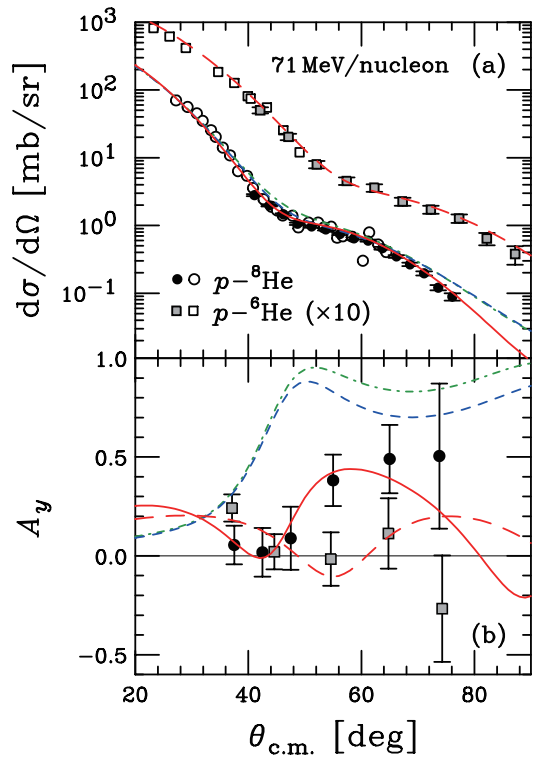}}
\caption{(Color online) The differential cross section (upper) and
analyzing power (lower) of $p$--$^{6,8}$He elastic scattering at
71~MeV/nucleon.} \label{a07_phen}
\end{figure}

The measured differential cross sections ($d\sigma/d\Omega$)
and analyzing powers ($A_y$) for
$p$--$^8$He (present) and $p$--$^6$He~\cite{Uesaka10,Sakaguchi11}
are shown in Fig.~\ref{a07_phen} as filled circles and squares,
respectively. Published $d\sigma/d\Omega$
data~\cite{Korsheninnikov97} are also plotted as the open symbols.
It is known from extensive measurements at 65~MeV~\cite{Sakaguchi82}
that the analyzing powers for $p$--$A$ scattering from stable nuclei
usually take large positive values of $\sim$0.9 at the second peak,
except for the $p$--$^4$He case in which $A_y$ is almost
zero~\cite{Burzynski89}. The present $A_y$ data for $p$--$^8$He and
$p$--$^6$He lie between these two cases.

To determine the spin-orbit potentials between a proton and $^8$He
nucleus, we perform a phenomenological optical model analysis. For
the optical model potential, we use a Woods-Saxon form factor with a
Thomas-type spin-orbit term:
\begin{eqnarray}
U_{\textrm{OM}}(R) =  &-&V_0 \, f_r (R) - i \, W_0 \, f_i (R)  \nonumber \\
      &+&  V_s \, \frac{2}{R} \, \frac{d}{dR} f_s (R) \; \bm{L} \cdot \bm{\sigma}_p
          + V_{\textrm{C}} (R) \label{eq1}
\end{eqnarray}
with
\begin{eqnarray} \label{eq2}
f_x (R) &=& \left[ 1 + \exp \left( \frac{R-r_{0x} A^{1/3}}{a_x}
                        \right) \right] ^{-1} \; \\
(x&=&r,i,\ \textrm{or}\ s).\nonumber
\end{eqnarray}
Here, $\bm{R}$ is the relative coordinate between a proton and a
${}^{8}$He particle,
$\bm{L} = \bm{R}\times(-i\hbar\bm{\nabla}_{R})$
is the associated angular momentum, and $\bm{\sigma}_{p}$ is the Pauli
spin operator of the proton.
The subscripts $r,\ i,\ \textrm{and}\ s$ denote the real and
imaginary parts of the central term and the real part of the
spin-orbit term, respectively. 
$V_0, W_0$ and $V_s$ are depth parameters of the corresponding terms.
$r_{0x}$ and $a_x$ are radius and diffuseness parameters, respectively.
$V_{\textrm{C}}$ is a Coulomb potential of uniformly charged
sphere with a radius of 
$r_{0\textrm{C}} A ^{1/3}$ fm ($r_{0\textrm{C}}=1.3$ fm).
No surface absorption term is considered here.
Since the statistics is limited, the imaginary part of the spin-orbit
potential is not included in the fits. If we assume it is as small as in
the case
of stable nuclei, the effect on $A_y$ is within the error bars.
However, as it is still unknown whether this assumption holds in
unstable nuclei,
the imaginary spin-orbit potential should be investigated in future when
sufficient data is available.

Using the optical potential given in Eqs.~(\ref{eq1}) and (\ref{eq2}),
we search for a parameter set that reproduces both the
$d\sigma/d\Omega$ and $A_y$ data obtained in the present work and
the $d\sigma/d\Omega$ data of Ref.~\cite{Korsheninnikov97}. The fit
is carried out using the {\sc ECIS79} code~\cite{Raynal65}. The
initial values are taken from a set of parameters for $p$--$^6$Li
elastic scattering at 72~MeV/nucleon~\cite{Henneck94}. The solid and
long-dashed curves in Fig.~\ref{a07_phen} show the best-fit results
for $p$--$^8$He and $p$--$^6$He, respectively. The reduced
chi-square values for $d\sigma/d\Omega$ and $A_y$ are minimized as
$\chi^2_{\sigma}/\nu_{\sigma}=$ 1.91 and $\chi^2_{A_y}/\nu_{A_y}=$
0.37, respectively, 
in the $p$--$^8$He case. 
The optical potential parameters of
$p$--$^6$Li~\cite{Henneck94} and $p$--$^6$He~\cite{Sakaguchi11} and
those obtained for $p$--$^8$He 
(Set-A)
are summarized in
Table~\ref{table:param}.
These three potentials are similar to each other, probably because of
the resemblance of density distribution.
Since $^6$Li is also a weakly-bound nucleus, 
its matter radius and $d\sigma/d\Omega$ are almost identical with those
for $^6$He as described in Ref.~\cite{Sakaguchi11}.
However, we should note that it is not straightforward to
deal with the spin-orbit potential for the $^6$Li case, because it has
a non-zero spin. Henceforth, the quantitative discussion focuses on the
nuclei with spin zero.

\begin{table*}[htbp]
\caption{Parameters of the optical potentials for $p$--$^6$Li at
72~MeV/nucleon~\cite{Henneck94}, $p$--$^6$He at
71~MeV/nucleon~\cite{Sakaguchi11}, and $p$--$^8$He at 71~MeV/nucleon
(present work).}
\begin{ruledtabular}
\begin{tabular}{c|ccccccccccc}
   &  $V_0$  &  $r_{0r}$  &  $a_{r}$  &  $W_0$  &  $r_{0i}$ &  $a_{i}$
   & $V_{s}$ & $r_{0s}$ & $a_{s}$ & $\chi^2_{\sigma}/\nu_{\sigma}$ & $\chi^2_{A_y}/\nu_{A_y}$ \\
   &  (MeV)  &  (fm)  &  (fm)  &  (MeV)  &  (fm)  &  (fm)
   &  (MeV)  &  (fm)  &  (fm) & & \\ \hline
 $p$--$^6$Li~\cite{Henneck94} & 31.67 & 1.10 & 0.75 & 14.14 & 1.15 & 0.56
 & 3.36 & 0.90 & 0.94 & & \\
 $p$--$^6$He~\cite{Sakaguchi11} & 27.86 & 1.074 & 0.681 & 16.58 & 0.86 & 0.735
 & 2.02 & 1.29 & 0.76 & 0.95 & 0.96 \\
 $p$--$^8$He (Set-A) 
 & 41.60 & 0.95 & 0.73 & 22.78 & 0.97 & 0.86
 & 3.68 & 1.11 & 0.91 & 1.91 & 0.37 \\
 $p$--$^8$He (Set-B) & 47.26 & 0.89 & 0.75 & 26.34 & 0.90 & 0.88
 & 4.15 & 1.06 & 0.95 & 2.40 & 0.34 \\
 $p$--$^8$He (Set-C) & 57.90 & 0.75 & 0.80 & 34.34 & 0.96 & 0.74
 & 2.65 & 1.17 & 0.86 & 1.93 & 0.25 \\
\end{tabular}
\end{ruledtabular}
\label{table:param}
\end{table*}

\begin{figure}[ht]
\centering
\resizebox{0.75\linewidth}{!}{\includegraphics{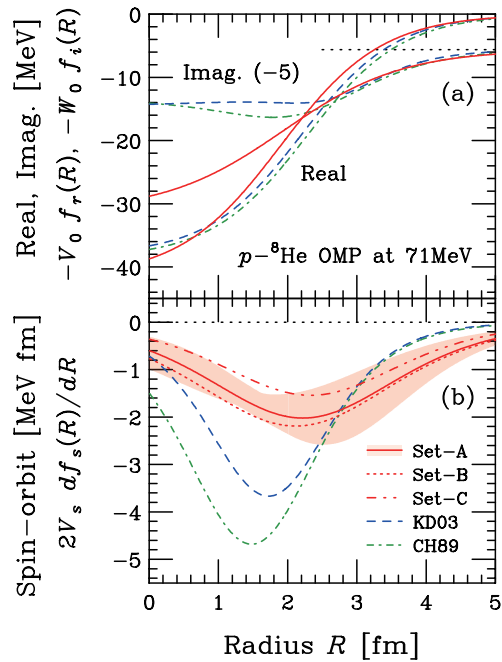}}
\caption{(Color online) Radial dependence of the optical potential between a proton and a $^8$He nucleus.}
\label{pcomp}
\end{figure}

The upper panel of Fig.~\ref{pcomp} presents the radial dependence
of the central terms of the $p$--$^8$He potential (Set-A).
The solid,
dashed, and dot-dashed curves denote the present potential, that
obtained by Koning and Delaroche (KD03)~\cite{Koning03}, and that
obtained by Varner~{\it{et al.}} (CH89)~\cite{Varner91},
respectively. A surface absorption term is included in the imaginary
term in the case of the global potentials. While the $^8$He nucleus
is located outside the applicable range of these two global
potentials, they serve as guides for comparison since their
mass-number dependence is not strong, especially for the spin-orbit
term. The real and imaginary terms of the present potential are in
reasonable agreement with the global potentials. The
root-mean-square (r.m.s.) radii and volume integral of each term are
summarized in Table~\ref{table:vol}.
The real and imaginary
terms of the present potential are comparable to those of the global
potentials.

\begin{table}[htbp]
\caption{Volume integral and r.m.s. radius of each term of the
 $p$--$^{6,8}$He potentials at 71~MeV/nucleon.}
\begin{ruledtabular}
\begin{tabular}{cc|ccc|ccc}
 & & $J_r/A$ &  $J_i/A$ &  $J_{ls}/A^{1/3}$ &  $\langle r_{r}^2 \rangle ^{1/2}$ &  $\langle r_{i}^2 \rangle ^{1/2}$ &  $\langle r_{ls}^2 \rangle ^{1/2}$ \\
 & & \multicolumn{3}{c|}{(MeV fm$^3$)} & \multicolumn{3}{c}{(fm)} \\
\hline
\multicolumn{1}{c|}{} & Ref.~\cite{Sakaguchi11} & 320 & 144 & 66$^{+24}_{-26}$ & 2.95 & 2.98 & 3.33$^{+0.23}_{-0.26}$ \\
\multicolumn{1}{c|}{$^6$He} & KD03 & 419 & 198 & 93 & 2.94 & 3.07 & 2.37 \\
\multicolumn{1}{c|}{} & CH89 & 466 & 232 & 108 & 3.01 & 3.25 & 2.29 \\
\hline
\multicolumn{1}{c|}{}  & Set-A & 371 & 261 & 107$^{+35}_{-41}$ & 3.08 & 3.52 & 3.58$^{+0.25}_{-0.20}$ \\
\multicolumn{1}{c|}{$^8$He} & KD03 & 413 & 191 & 95 & 3.04 & 3.22 & 2.52 \\
\multicolumn{1}{c|}{}  & CH89 & 455 & 235 & 114 & 3.11 & 3.40 & 2.44 \\
\end{tabular}
\end{ruledtabular}
\label{table:vol}
\end{table}

The lower panel of Fig.~\ref{pcomp} displays the radial dependence
of $R V_{ls} (R)$, which is defined as
\begin{equation}
R V_{ls} (R) = 2 V_s \frac{d}{dR} \left[ 1 + \exp \left( \frac{R-r_{0s} A^{1/3}}{a_s}
                        \right) \right] ^{-1}.
\end{equation}
Here, the $R$ factor on the left-hand side is introduced to cancel
the $1/R$ term of the Thomas function in order to present the shape
of potential without divergence at small radii. The solid line in
Fig.~\ref{pcomp} (lower) shows the 
best-fit potential (Set-A) with a 
statistical error band (shaded area)
corresponding to a potential with $\Delta \chi^2_{A_y} \equiv
\chi^2_{A_y}-\chi^2_{A_y: {\textrm {min.}}}=1$. 
To check the fitting ambiguity of the spin-orbit potential, we search for
other possible parameter sets.
Excluding very unusual potentials such as ones with $V_0>$ 60 MeV, ten
different sets are obtained.
In Table~\ref{table:param}, two of them are presented:
Set-B and Set-C are the results with the deepest and the
shallowest spin-orbit potentials, respectively.
They are approximately consistent with that of Set-A within the
statistical error
band as shown in the lower panel of Fig.~\ref{pcomp}.
The obtained spin-orbit potentials have broad peaks at $R\sim2.2$~fm,
whereas
the global potentials (dashed and dot-dashed) have sharper peaks at
smaller radii of $R\sim1.6$~fm.
The spin-orbit potential for $^8$He
is found to be shallower and more
diffuse than the global systematics of stable nuclei.

To examine the effect of spin-orbit potential on the observables,
we compare the results of calculations using different spin-orbit
potentials but with identical central potentials. The short-dashed
and dot-dashed lines in Fig.~\ref{a07_phen} correspond to the
results of calculations using the same central terms as the present
potential but with the spin-orbit terms of the KD03 and CH89
potentials, respectively. These ``standard'' spin-orbit potentials
give large positive $A_y$ values that are incompatible with the
current data. It should be stressed that the shallow and diffuse
spin-orbit potential is essential in reproducing the present $A_y$
data.

\begin{figure}[ht]
\centering
\resizebox{0.8\linewidth}{!}{\includegraphics{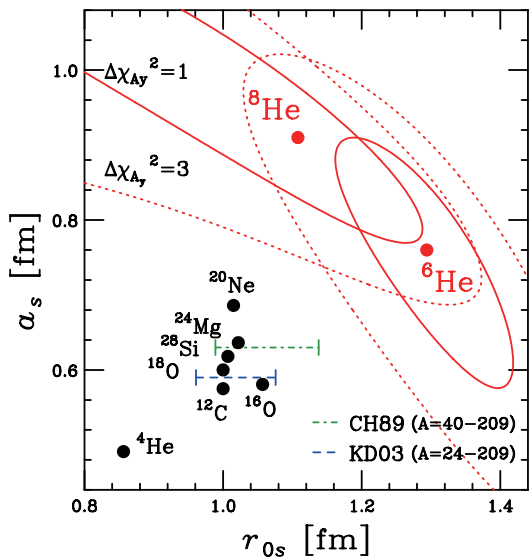}}
\caption{(Color online) Two-dimensional distribution of the radius
and diffuseness parameters of the spin-orbit term of the local
(filled circles) and global (dot-dashed: CH89, dashed: KD03)
potentials for spin-zero nuclei. 
Solid and dotted lines indicate $\Delta \chi ^2_{A_y}=$
1 and 3, respectively. } \label{fig:ct2-11}
\end{figure}

In Fig.~\ref{fig:ct2-11}, the parameters $r_{0s}$ and $a_s$ are
presented for comparison. Filled
circles show the parameters determined for spin-zero nuclei
ranging from $^4$He to
$^{28}$Si~\cite{Sakaguchi82,Fabrici80,Sakaguchi11}.
Parameters for heavier nuclei are represented by the global
potentials, KD03 (dashed) and CH89 (dot-dashed), which overlap those
of the light nuclei. 
The present results 
(Set-A for $^8$He)
are shown by the filled red
circles with uncertainties evaluated in the following manner: For
each point in the $r_{0s}$--$a_s$ plane, a depth parameter $V_s$ is
re-searched to minimize the $\chi^2_{A_y}$ value. The solid and
dotted lines in the figure indicate regions where $\Delta
\chi^2_{A_y}=$ 1 and 3, respectively. The radius and diffuseness
parameters of the spin-orbit potentials obtained for the
neutron-rich helium isotopes appear to be larger than those for the
stable nuclei. In contrast, the depth parameters for $^6$He and
$^8$He, determined as 2.02$^{+0.82}_{-0.86}$~MeV and
3.68$^{+0.80}_{-0.91}$~MeV, respectively, are smaller than the
typical value of $\sim$5~MeV.

\begin{figure}[ht]
\centering
\resizebox{0.9\linewidth}{!}{\includegraphics{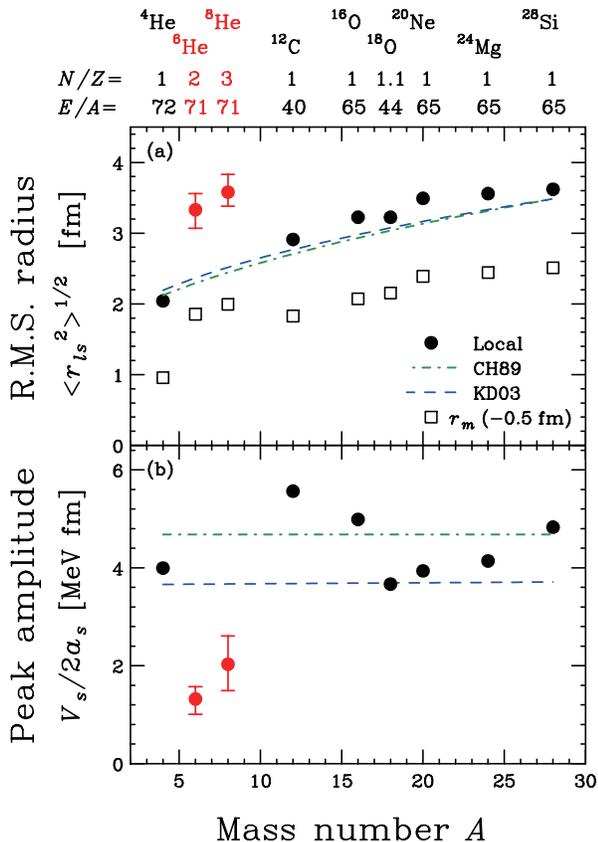}}
\caption{(Color online) The $\langle r_{ls}^2 \rangle ^{1/2}$ and
$r_m$ (upper) values and the peak amplitude $V_{s}/2a_{s}$ (lower)
of the spin-orbit potentials for light spin-zero nuclei. 
For $^8$He, the results of Set-A are shown. Those with Sets-B and
 -C are consistent with them within the statistical uncertainties.
The symbols for $r_m$ are shifted horizontally by $-$0.5~fm to prevent
 overlap.
}
\label{fig:lsall}
\end{figure}

The shape and magnitude of the spin-orbit potential can be
discussed in terms of the r.m.s. radius $\langle r_{ls} ^2 \rangle
^{1/2}$ and the amplitude of $R V_{ls}(R)$ at the peak position.
These quantities provide more robust features of the spin-orbit
potentials than the individual parameters that couple with each
other. 
Figure~\ref{fig:lsall}(a) shows the mass-number dependence of
the $\langle r_{ls} ^2 \rangle ^{1/2}$ values of the potentials for
the spin-zero nuclei. The symbols are the same as those in
Fig.~\ref{fig:ct2-11}. We can see that the $\langle r_{ls} ^2
\rangle ^{1/2}$ values of the present potentials (in red;
3.33$^{+0.23}_{-0.26}$~fm for $^6$He and 3.58$^{+0.25}_{-0.20}$~fm
for $^8$He) are significantly larger than the systematics of the
stable nuclei. Moreover, it is interesting to note that a close
similarity is found between the mass-number dependence of $\langle
r_{ls} ^2 \rangle ^{1/2}$ and that of the matter radius
$r_m$~\cite{Angeli04,Tanihata92,Alkhazov04,Kiselev05}, plotted as
the open squares in Fig.~\ref{fig:lsall}(a). The enhancement seen in
the $r_m$ values of $^6$He and $^8$He is more distinct in the
behavior of the $\langle r_{ls}^2 \rangle ^{1/2}$ values, which
indicates the particular sensitivity of the spin-orbit interaction to
the nuclear surface structure.

Figure~\ref{fig:lsall}(b) displays the amplitude of $RV_{ls}(R)$ at
the peak position $R=r_{0s}A^{1/3}$, which is denoted by $V_s/2a_s$.
The peak amplitudes of the local potentials for a stable nuclei are
in the range 3.5--5.5~MeV~fm and are almost independent of the mass
number. Those of the global potentials (dashed and dot-dashed lines)
are consistent with these amplitudes. However, the peak depths of
the present potentials, 1.32$^{+0.25}_{-0.21}$~MeV fm for $^6$He and
2.03$^{+0.58}_{-0.54}$~MeV fm for $^8$He, are smaller than the
standard values. From these results, we can conclude that the
spin-orbit potentials between a proton and neutron-rich $^6$He and
$^8$He nuclei are both shallower and more diffuse than the global
systematics of nuclei along the stability line. This is considered
to be a consequence of the diffuse density distribution of these
neutron-rich isotopes.

In summary, vector analyzing powers have been measured for the
proton elastic scattering from $^8$He at 71~MeV/nucleon to
investigate the spin-orbit potential between a proton and a
neutron-rich $^8$He nucleus. The measured differential cross
sections and analyzing powers were analyzed using a phenomenological
optical model to derive the overall characteristics of the
$p$--$^{6,8}$He interactions. The spin-orbit potentials for $^6$He
and $^8$He were found to be both shallower and more diffuse than the
global systematics of stable nuclei. The r.m.s. radius of these
spin-orbit potentials deviate from the well-established mass-number
dependence and show a close similarity to the behavior of the matter
radius. Depths of the obtained potentials were found to be
significantly reduced from the standard value. The shallow and
diffuse spin-orbit potentials for $^6$He and $^8$He are considered
to be a consequence of the diffuse density distribution of these two
neutron-rich helium isotopes.

The authors thank the staff at the RIKEN Nishina Center and CNS for
operating
the accelerators and ion source during the measurement. 
We acknowledge Y.~Iseri, M.~Tanifuji, and S.~Ishikawa for 
fruitful discussions.
This work was supported by MEXT KAKENHI Grant Number 17684005.
S.~S. acknowledges financial support from JSPS KAKENHI Grant Number
18-11398.

\bibliographystyle{prl}

\end{document}